\documentstyle[seceq,epsf]{ptptex}

\title{
Stability of Relativistic Blast Waves 
}

\author{
Jun {\sc Ogura}\footnote{E-mail:
ogura@theo.phys.sci.hiroshima-u.ac.jp}  
and Yasufumi {\sc Kojima}\footnote{E-mail:
kojima@theo.phys.sci.hiroshima-u.ac.jp} 
}

\inst{
Department of Physics, Hiroshima University, Hiroshima 739--8526, Japan}

\recdate{
November 12, 1999
}
\notypesetlogo

\abst{
A spherical blast wave with relativistic velocity
can be described by a similarity solution, that is
used for theoretical models of gamma-ray bursts.
We consider the linear stability of such a relativistic blast wave
propagating into a medium with density gradient.
The perturbation can also be expressed by a
self-similar form. We show that 
the shock front is  unstable in general, and
we evaluate the growth rate.
}

\begin{document}

\maketitle

\section{Introduction}

  Gamma-ray bursts have puzzled astronomers since their 
accidental discovery in the late sixties. Recent observations still
provide new topics of study, afterglow,\cite{Cosra} optical
flash,\cite{Akerlof}
brightest events,\cite{Castro} and so forth.  
In these bursts, huge amounts of energy are released instantaneously.
The  subsequent evolution of the radiation-dominated matter can be 
described by an expanding blast wave.
Gamma-rays are emitted when a matter with highly 
relativistic velocity, with Lorentz factor 
$\Gamma \geq 10^2 $, is converted into radiation. 

The observed fluctuations in gamma-rays with a short timescale
are thought to be associated with the complex 
behavior of shock waves. Two locations have been proposed as the origin 
of the irregularities.
One location is inside the relativistic fluid, as described by the
internal shock 
model. 
When shock waves intersect each
other, gamma-rays are emitted. The complexity of the corresponding
time profile  depends on  the internal shock structure.
The other location is in the shock front itself, as described by
the external shock model.  
When the shock front impinges upon ISM (inter stellar matter), 
gamma-rays are emitted. 
Both models have advantages and disadvantages. 
Waxman\cite{waxman-a}\cite{waxman-b} showed synchrotron
emission from external shocks provides a successful model for the
broken power law spectra and smooth temporal behavior of afterglow. 
Sari and Piran,\cite{Sari1} however, showed that external shocks cannot
produce variable gamma-ray bursts unless they are produced by 
extremely narrow jets or if 
only a small fraction of the shell emits radiation.
The expanding shell and ISM were assumed to be smooth.
The observed nature with rapid variability
depends on the fluctuations of the external matter,
which is not yet known.
The possibility of irregularity in the shock front can be investigated.
An irregular shell with angular fluctuations in the external shock may
produce temporal behavior of the gamma-ray emission. 

  A theoretical model of the blast wave with ultra-relativistic 
speed was considered by Blandford and McKee.\cite{Blandford} They 
obtained a spherical  self-similar solution, which can 
be regarded as the relativistic version of the Sedov-Taylor solution. 
Using a numerical code for the spherical symmetry,
Kobayashi, Piran and Sari\cite{Sari2} simulated
an expanding shock with relativistic speed and confirmed
that the self-similar solution is a good approximation.
The spherical symmetry may be too great a simplification in some cases. 
Indeed, in non-relativistic hydrodynamics, 
Ryu and Vishniac\cite{Ryu} showed that under certain conditions 
the spherical shock becomes overstable with respect to perturbations
with tangential velocity. From this point of view, it is important
to examine the dynamical stability of the Blandford-McKee solution.

In this paper, we consider the linear stability with respect to
non-spherical deformations of ultra-relativistic shock waves. 
There is a similarity solution describing shocks with an impulsive
energy supply.  
In Section 2,  we show that the perturbation functions can 
be obtained in a power low form.
Matching the boundary conditions at the shock front 
and using a regularity condition at the origin,
the temporal behavior, i.e. an eigenvalue of the functions,
is determined.
Section 3 is devoted to discussion.

\section{Basic equations and approximate solutions}

\subsection{Sphrical shock waves}

Relativistic fluid motion is described by 
conservation laws of number and energy-momentum.
We explicitly give these equations in spherical 
coordinates for two-dimensional flow with the velocity 
$(v_r , v_\theta)$ and the Lorentz factor $ \gamma  $
as
\begin{subequations}

\begin{eqnarray}
\partial_t\,D + \frac{1}{r^2}\,\partial_r\,(r^2\,D\,v_r)
    + \frac{1}{r\,\sin\theta}\partial_\theta\,(\sin\theta\,D\,v_\theta)&=& 0,
    \label{1-def.eq.1}\\
  \partial_t\,(W - p) + \frac{1}{r^2}\,\partial_r\,(r^2\,W\,v_r)
  + \frac{1}{r\,\sin\theta}\,\partial_\theta\,(\sin\theta\,W\,v_\theta) &=& 0,\\
 \partial_t\,(W\,v_r) +
  \frac{1}{r^2}\,\partial_r\,\{r^2\, (W\,v_r ^2 + p)\,\}
 +\frac{1}{r\,\sin\theta}\,\partial_\theta\,(\sin\theta\,W\,v_r\,v_\theta) \nonumber\\ 
  -\frac{W}{r}\,v_\theta ^2 - \frac{2 p}{r} &=& 0,\\
 \partial_t(W\,v_\theta)
      + \frac{1}{r}\,\partial_r\,(r\,W\,v_r\,v_\theta)
      + \frac{1}{r\,\sin\theta}\,\partial_\theta\,
      \{\sin\theta\,(W\,v_\theta ^2 + p)\,\} \nonumber\\
      + \frac{2}{r}\,W\,v_r\,v_\theta - \frac{p}{r}\,\cot\theta &=&0,
      \label{1-def.eq.4} 
\end{eqnarray}
\end{subequations}
where 
\begin{eqnarray}
 D &\equiv& n\,\gamma , \\
 W &\equiv & (e + p)\,\gamma^2 = 4p\gamma^2 ,  \label{2.1-def.of.enthalpy-W}
\end{eqnarray}
and $ n,  e$  and $ p$ are the number density, energy density and
pressure. We have assumed that after a strong shock, that the
relativistic fluid is radiation dominated, i.e. $  e = 3p$.

Blandford and McKee\cite{Blandford} obtained a spherically symmetric
similarity solution of Eqs.~(\ref{1-def.eq.1}) -- (\ref{1-def.eq.4}). 
This solution can be written in terms of $t $ and $r$. 
The kinematical  energy of the shock decreases
with time. The Lorentz factor $\Gamma$ 
of the shock front therefore indicates the timescale. 
In place of the radial coordinate, they introduced 
a new coordinate $\chi$ as the similarity variable.
The position of the shock front corresponds to $ \chi = 1 ,$
and the interior  region  corresponds to $ \chi > 1 .$
These new coordinates are mathematically defined as  
\begin{eqnarray}
 \Gamma^2 &=& (t/t_0)^{-m}, \label{2.1-def-of-Gamma} \\
 \chi &=& \{1 + 2\,(m + 1)\,\Gamma^2\}\,(1 - r/t),
\end{eqnarray}
where $t_0 $ is a constant, and the index $m$ determines 
the energy supply rate of the blast wave.
For the $m = 3 $ case, the motion corresponds to an adiabatic point 
explosion, that is, an impulsive injection of energy at the center.
For the case $ m<3 $, the decrease of the velocity is less than that for
the adiabatic case.  This means
that the solution corresponds to a blast wave with
additional power supply.\cite{Blandford}
This solution has additional inner shock and a contact 
discontinuity inside the flow.

We assume a strong shock, that is, that the kinematical shock energy is
extremely large. This situation 
corresponds to a large value of $\Gamma .$
The solution matched with the strong  shock condition
can be expressed  by
\begin{eqnarray}
 D_0 &=& 2\,n_1\,{\Gamma}^2\,h \ , \label{2.3-new.def-D} \\
 p_0 &=& \frac{2}{3}\,w_1\,{\Gamma ^2}\,f \ , \label{2.3-new.def-p}\\
 W_0 &=& \frac{4}{3}\,w_1\,\Gamma^4\,f\,g \ ,\\
( v_{r0}, \, v_{\theta 0} ) &=& 
\left( 1 - \frac{1}{\Gamma^2\,g}  ,\, 0 \right)\ ,\\ 
 \gamma ^2 &=& \frac{1}{2}\,{\Gamma}^2\,g \ ,
  \label{2.3-new.def-gamma} 
\end{eqnarray}
where $n_1$ and $w_1$ are the number density and enthalpy
in front of the shock.
Since we consider a strong shock, 
the external medium is cold and non-relativistic, and thus the pressure
of the medium in front of the shock can be neglected.
We also assume that the density $n_1$ varies 
as a power of the radius:
\begin{eqnarray}
 n_1 \propto r^{-k}.
\end{eqnarray}
The functions $f(\chi)$, $g(\chi)$ and $ h(\chi) $ can be  
obtained by solving differential equations, but they reduce to 
simple analytic forms for the special case $m=3-k $
as
\begin{subequations}
\begin{eqnarray}
f &=& \chi ^{- (4k - 17)/(3k - 12)} ,\\
g &=& \chi ^{- 1} ,\\
h &=& \chi ^{- (2k - 7)/(k - 4)} .
  \label{2.1-h-chi}
\end{eqnarray}
\end{subequations}
We only examine the stability of the case that includes the point
explosion into the constant density $k=0$.
The calculation is simply, and the result is likely to be
common to other cases,
in which the growth/decay  rate may be modified.

\subsection{Perturbation equations}

We denote the Eulerian variation of a variable by $\delta $, 
and we expand variables, e.g. the pressure as
\begin{equation} 
 p= p_0 \,+\,\delta\,p =  
          p_0\, \{ 1 + (\Gamma ^2)^{s}\,
	                    \xi_P\,P_l\} \ ,
\end{equation} 
where $ P_l(\cos \theta)$ is the Legendre polynomial of order $l $,
and a power law form is assumed with respect to
the ``time'' $\Gamma .$ 
Since the background quantities change in a self-similar way,
the perturbations also have self-similar forms.
The index  $s$ is a complex number in general. The real part of it 
represents the growth rate for $ \Re(s) <0, $
since the factor $ \Gamma $ decreases with time. 
%
In a similar way, perturbations of the density and velocity are expanded
as  
\begin{eqnarray}
D &=&
 D _0 +\,\delta\,D\, =
          D_0\, \{ 1 + (\Gamma ^2)^{s}\,\xi_D \,P_l\} \ , \\
v_r&=&  v_{r 0}\,\,+\,\delta\,v_r  =  
          1 - \frac{1}{\Gamma^2\,g}\,
	   \{1 + (\Gamma ^2)^{s}\,\xi_R \,P_l\} \ ,\\
v_{\theta} &=&
 \delta\,v_{\theta 0} =  
                (\Gamma ^2)^{s}
		     \,\xi_T\,\frac{d}{d\,\theta}P_l \ .
\end{eqnarray}

  A perturbation of the function $ W$ can be expressed 
by the perturbations of the pressure and radial velocity as
\begin{equation}
 W= W_0 \,+\,\delta\,W   =  
          W_0\, \{ 1 + (\Gamma ^2)^{s}\,g\,
                   (2\,\xi_P + \xi_R) P_l\} \ .
\end{equation}
Using the definitions in Eqs.~(\ref{1-def.eq.1}) -- (\ref{1-def.eq.4}),
we have the linearized  perturbation equations.
After separating the time and angular parts, we have 
the differential equations for the functions
$\xi_D, \xi_P, \xi_R $ and $\xi_T .$
The explicit forms can be written as
\begin{subequations}

\begin{eqnarray}
(k - 4) \chi \frac{d \xi_D}{d \chi}
  - \frac{3}{2} (k - 4) \chi\frac{d \xi_P}{d \chi}
  &=&
 (k - 3) s \xi_D + \frac{4}{3} (k -2) \xi_R \nonumber\\
 && - \frac{1}{2} \left\{- 3 ( s + 4) + k (s + 6)\right\}\xi_P ,
         \label{org-4.self.perturb.eq.1} \\
  (k - 4)\chi \frac{d \xi_R}{d \chi}
  + \frac{3}{2} (k - 4) \chi \frac{d \xi_P}{d \chi}
  &=&
  - \left\{ 3s + \frac{5}{3}
   - k \left(s + \frac{1}{3}\right)\right\} \xi_R \nonumber\\
 &&{} + \frac{1}{2}\left\{ 3(s-4) - k(s-6)\right\} \xi_P ,
  \label{org-4.self.perturb.eq.2}   \\
 (k -4) \chi \frac{d \xi_T}{d \chi}
  &=&
 - \left\{ 3 s + \frac{14}{3}
    - k \left(s + \frac{7}{3}\right)\right\} \xi_T ,
  \label{org-4.self.perturb.eq.3}   \\        
 (k -4) \chi \frac{d \xi_D}{d \chi} + 2(k - 4)\chi \frac{d \xi_R}{d \chi}
  &=&
  (k - 3)s \xi_D - 2(3 k - 7) \xi_R \nonumber\\
     &&- l (l+ 1) \xi_T ,
  \label{org-4.self.perturb.eq.4} 
 \end{eqnarray}
\end{subequations} 
where we have retained the highest order of $\Gamma$ only.
The differential Equations~(\ref{org-4.self.perturb.eq.1}) --
(\ref{org-4.self.perturb.eq.4}) can be
solved analytic ally.
The solutions are expressed with four integration constants,
$a, b, c$ and $d$:
\begin{subequations}

\begin{eqnarray}
 \xi _D &=& 
  a \chi ^{p_1} + 
   b q_1\,l\,(l +1)  \chi ^{p_2} +
    c q _{4_-} \chi ^{p_+} +
     d q _{4_+} \chi ^{p_-} \ , \label{ogr-5-last.m3.perturb.eq.1}\\
 \xi _R  &= & 
   b q_2\,l\,(l +1)  \chi ^{p_2 } +
    c q _{5_+} \chi ^{p_+} +
     d q _{5_-} \chi ^{p_-} \ , \label{ogr-5-last.m3.perturb.eq.2}\\
 \xi _P  &=& 
   b q_3\,l\,(l +1)  \chi ^{p_2 } +
    c  \chi ^{p_+} +
     d  \chi ^{p_-} \ , \label{ogr-5-last.m3.perturb.eq.3}\\
 \xi _T &=&  
  b   \chi ^{p_2 } \ . \label{ogr-5-last.m3.perturb.eq.4}
\end{eqnarray}
\end{subequations}
Here the indices $ p_i $ and coefficients $q_i $ depend on $ s$ only:
\begin{subequations}
\begin{eqnarray}
p _1 & = & \frac{k-3}{k-4}\,s \ ,\\
 p _2 & = &\frac{ k (3s + 7)-9s - 14 }{3(k - 4)} \ ,\\
 p _\pm & = &
  \frac{- k(6s + 17)+18s + 43  \pm f_1}{6(k-4)} \ ,
\end{eqnarray}
\end{subequations}
where
\begin{eqnarray}
 f_1 &=& \left\{  {k^2}\,\left(  48\,{s^2} + 192\,s +841 \right)
      -2\,k\,\left(144\,{s^2}+528\,s + 1943  \right)\right.\nonumber\\
        &&{}\left.+ 432\,{s^2}+ 1440\,s + 4489   \right\}^{1/2}
\end{eqnarray}
and
\begin{subequations}

\begin{eqnarray}
q _1 & = &
  - \left[7\,\left( k - 2 \right) \,
  \left\{  2\,{k^2}\,\left( 4\,{s^2}+23\,s +   5 \right)
   -   k\,\left(  48\,{s^2}+ 240\,s +43   \right) \right.\right.\nonumber\\
 &&{} \left.\left. +  72\,{s^2}+ 306\,s + 46    \right\}    \right]^{-1}
\left\{ {k^2}\,\left( 4\,s  - 2\right)
      -  k\,\left( 26\,s - 5 \right) + 42\,s -2  \right\} \ ,\\
 q _2 & = &  - \left\{
  2\,{k^2}\,\left( 4\,{s^2} + 23\,s +  5  \right)
   -   k\,\left(48\,{s^2}+ 240\,s +  43  \right)
   + 72\,{s^2} + 306\,s + 46\right\}^{-1}\nonumber\\
  &&{}\times \bigl\{ k(4s + 1) -2(6s + 1) \bigr\},\\
 q _3 & = & 
\left\{
  2\,{k^2}\,\left( 4\,{s^2} + 23\,s +  5  \right)
   -   k\,\left(48\,{s^2}+ 240\,s +  43  \right)
   + 72\,{s^2} + 306\,s + 46 \right\}^{-1}\nonumber \\
  &&{} \times\bigl\{2( 2k - 3 )\bigr\},\\
 q _{4_\pm} & = & \left[4\,
          \bigl\{ k\,\left(  s + 4 \right) - 3\,s -10   \bigr\}
          \left\{{k^2}\,\left( 4\,{s^2}+ 9\,s   -23  \right)
           - k\,\left( 24\,{s^2}+ 50\,s  - 101 \right)
	   \,\right.\right.\nonumber \\	       
         &&{} \left.\left.      + 36\,{s^2} + 69\,s -110 
                    \right\}\right]^{-1}
         \left[ {k^3}\,\left( 12\,{s^3} + 113\,{s^2}+ 231\,s
	  -406 \right) \right.\nonumber\\
         &&\left.-{k^2}\,\left(108\,{s^3}+ 949\,{s^2}
		+ 1822\,s-2736\right)
            + k\left(324\,{s^3}+ 2643{s^2}+ 4763s-6126\right)
            \right.\nonumber\\
             &&{}\left.	    
            - 324\,{s^3}- 2439\,{s^2}- 4128\,s  + 4556
            \pm  \left\{ {k^2}\,\left( {s^2} + 3\,s -14 \right)
            \right.\right.\nonumber\\
            &&{}\left.\left.
                  - k\,   \left(  6\,{s^2}+ 17\,s  -62  \right)  
                  + 9\,{s^2}+ 24\,s -68
                  \right\}  f_1         \right] , \\
 q _{5 _ \pm} & = &
                  \bigl[8\,\bigl\{ k\,\left( s  + 4\right)
	                  - 3\,s -10    \bigr\}\bigr]^{-1}
		   \bigl(-29k + 67 \pm f_1\bigr) .
\end{eqnarray}
\end{subequations}
The physical interpretation of the solutions is evident. 
The term with the constant $a$ simply represents
the perturbation of the number density, without
disturbing any other quantities.
The terms with the constant $b $ represent
the perturbation of the angular momentum.
This mode is possible only for non-spherical 
perturbations $(l \ne 0).$
For a spherically symmetric  perturbation, $\xi _T =0$, $l = 0$. In
this case, 
the solution should be reduced by setting $ b =0 $ in
Eqs.~(\ref{org-4.self.perturb.eq.1}) -- (\ref{org-4.self.perturb.eq.4}).
The terms involving the constants $c$ and $d$ represent
the pressure waves propagating in the inward and outward 
directions.

\subsection{Regularity condition at the origin}

  For the boundary condition at the origin, we require that
the fluid does not undergo divergent perturbations.
Also, the pressure perturbation should vanish:
$ \delta p \rightarrow 0 \ .$
This condition corresponds to 
\begin{equation}
\xi_P \rightarrow 0 
\end{equation}
for $\chi \rightarrow \infty$.
It restricts the possible form of the power index $\Re(s)$; that is, 
$\xi_P$ must decrease as the origin is approached.
The explicit condition is written as
\begin{equation}
 \Re(s) < \frac{7(k+2)}{3(k-3)} . \label{bound-R}
\end{equation}

\subsection{Boundary conditions at the shock front} \label{3.2-section}

Here we consider the deformation of an expanding shock wave. 
The radius of the shock is assumed to be given by
\begin{equation}
 r =  t \left\{ 1- \frac{1}{ 2(m+1) \Gamma ^2 } \right\}  + \eta ,
\end{equation}
where  the first term denotes the radius of the unperturbed shock 
and the second denotes the perturbation.
We also assume that the perturbation function has the power-law form
\begin{equation}
\eta = - \frac{\alpha}{2}\,t\,\left(\Gamma^2\right)^{s-1} 
P_l \ , \label{3.1-def.of.eta}
\end{equation}
where $\alpha$ is the normalization factor. 
This change of the shock front induces a radial velocity as
\begin{equation}
  \delta V_{r} = 
  \frac{d \eta}{d t} = - \alpha \frac{4 - 3s}{2}\left(\Gamma ^2\right)^{s-1}
  P_l \ .
\end{equation}
We now apply the jump condition for
the perturbation functions at this perturbed shock front.
That is, we stipulate that the number flux, energy  flux  and momentum 
flux be continuous  across the surface (see \ref{app.AA}).
The values at the perturbed position are
evaluated by the Lagrange perturbation. For example, 
the pressure at the perturbed shock position is given by 
\begin{eqnarray}
\Delta p = \delta p  + \eta \frac{\partial p_0}{\partial r}
 \ . \label{3.1-def.of.DELTA} 
\end{eqnarray}

It is also easy to rewrite the boundary conditions
at shock front as
\begin{subequations}
\begin{eqnarray}
  \xi_D / \alpha &=& 21 s - 10 k +  13,
             \label{org-6.2-self.m3.bound.cond.at.shock.1}\\   
 \xi_P / \alpha &=& 9 s - 4 k + 5,
             \label{org-6.2-self.m3.bound.cond.at.shock.2}\\  
 \xi_R / \alpha &=& 12 s - 5 k + 5 .
               \label{org-6.2-self.m3.bound.cond.at.shock.3}
\end{eqnarray}
\end{subequations}
\subsection{Approximate solution} \label{section-2.5}

\begin{figure}[htbp]
  \begin{center}
    \leavevmode
    \epsfxsize=12cm \epsfbox{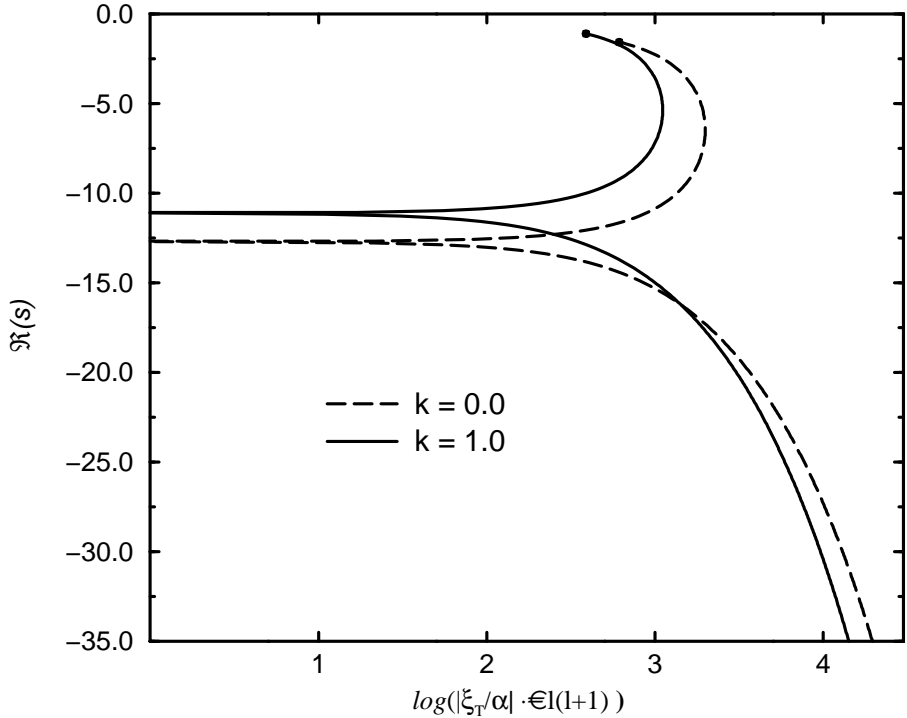}
    \caption{The growth rate as a function of normalized tangential
   velocity, $|\xi_T   / \alpha| \cdot l (l + 1).$ Two curves are shown
   corresponding to the external density distribution $k$.  Uniform case
   is $k=0.$ Dots near $\Re(s) =-1.5 $ are the maximum values which are
   determined from the regularity condition at the origin.} 
    \label{graph1}
  \end{center} 
\end{figure}

We can easily determine the integration constants
by matching the jump conditions to the interior solution 
Eqs.~(\ref{ogr-5-last.m3.perturb.eq.1}) -- (\ref{ogr-5-last.m3.perturb.eq.4}) 
(see \ref{appD}).
The temporal dependence $s$ is specified by one parameter,
the velocity in the tangential direction.
We use the normalized value 
$|\xi_T/\alpha| \cdot l (l + 1)$
and plot the growth rate $ \Re(s) $ as a function of it in Fig.~1. 
Both the cases $\xi_T/\alpha < 0$ and  $\xi_T/\alpha > 0 $
are shown. 
For the spherically symmetric case
$ \xi_T/\alpha \cdot l (l + 1) =0$,
the growth rate is  $ \Re(s) \sim - 12.7$ for $k = 0$.
As the value  $\xi_T/\alpha $ decreases, 
$ \Re(s) $ increases. 
We only show the region of $\Re(s)$ limited by Eq.~(\ref{bound-R});
that is, the curves are terminated near  $ \Re(s) = -1.5.$
As the value  $\xi_T/\alpha $ increases from zero, 
$ \Re(s) $ monotonically decreases.
This implies that the temporal behavior depends significantly 
on the $\Gamma $ factor. The mode grows strongly with the deceleration
of the background fluid.
In particular, the growth rate is large 
for shorter wavelengths in the tangential direction. 
In this figure, we also show the effect of the external density
gradient,
uniform ($k=0$) and a decreasing slope with the radius ($k=1$). 
The general features of the growth rate 
do not depend on the density distribution.
The growth rate is  modified by about 10\%.

\begin{figure}[htbp]
  \begin{center}
    \leavevmode
    \epsfxsize=12cm \epsfbox{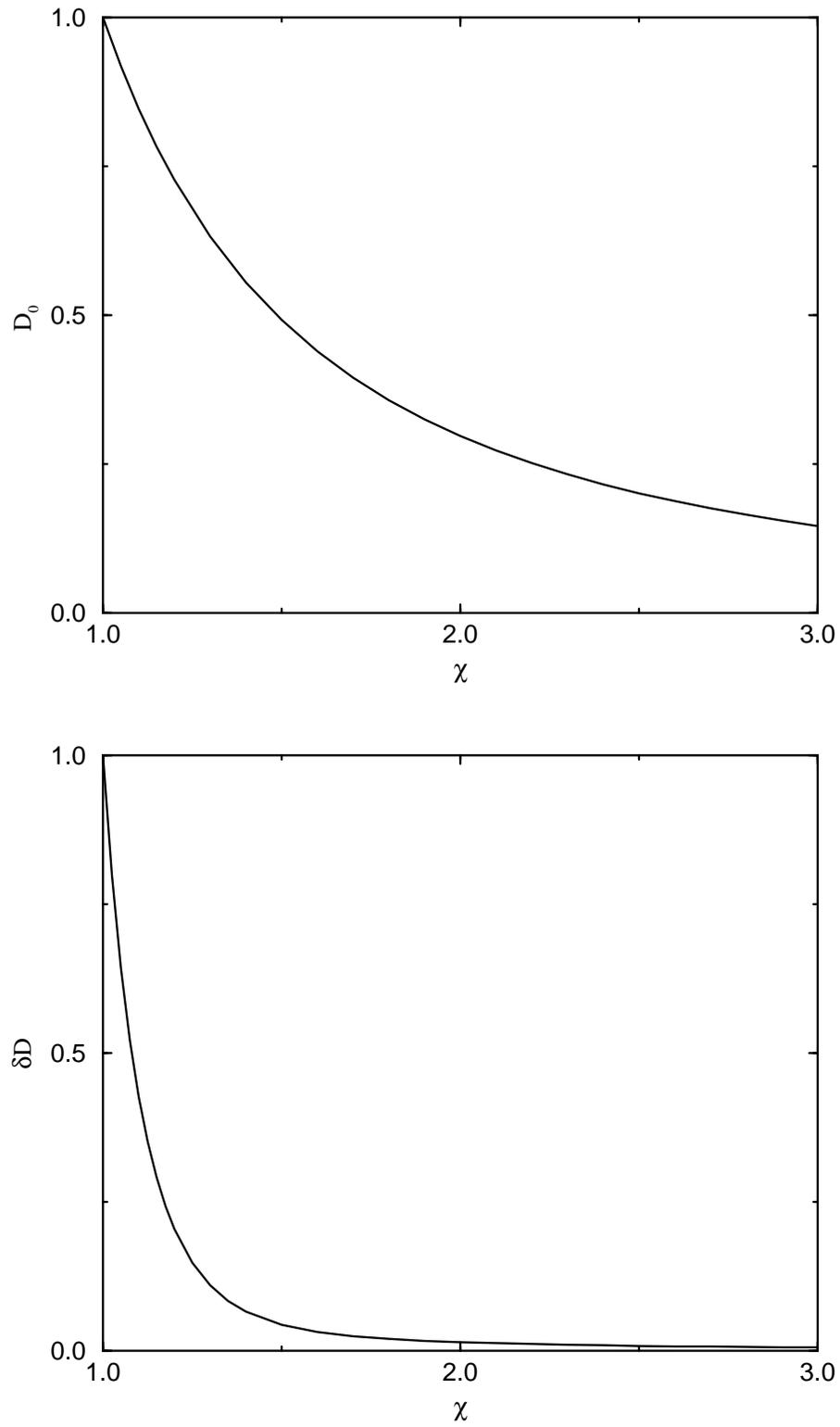}
    \caption{The shape of background density($D_0$) and its
   perturbation value($\delta D$) normalized at shock front. The
   parameters adopted here are $B l(l+1) = 1000$ and $k = 0$. $\chi$ is
   a similarity variable 
   which show 
   the distance from shock front, where $\chi = 1$ is the shock front.}
    \label{graph2}
  \end{center} 
\end{figure}

The radial functions of the background and perturbation fields 
are shown in Fig.~2.
We display the density profiles near the shock front,
which corresponds to $\chi =1 .$ 
The thickness of the expanding shell is approximated by the region 
from $\chi =$ 1  to  2.
The radial function of the perturbation is localized 
within the small region of the shell; that is,
it is a quite steep function.


\section{Summary and discussion}

In this paper, we have studied the linear stability of 
a spherically expanding shock in the ultra-relativistic limit,
i.e. the large $ \Gamma $ limit.
The temporal behavior of the perturbation
is expressed by a power law in $ \Gamma$.
We found a solution that grows with the deceleration of the shock.
The growth rate becomes large for short wavelengths of the angular
fluctuation. That is, there exists a rapidly growing 
mode for the non-spherically symmetric displacement.
%
Vishniac\cite{Vishniac} discussed the physical interpretation of the 
unstable mode in the non-relativistic blast wave.
It is important to remember that two kinds of pressures 
exerted to the shock plane have completely different natures.
In the exterior region before the shock,
the ram pressure dominates and is parallel to the direction of the shock
propagation.
After the shock, the thermal pressure dominates and is isotropic. 
When the shock spherically expands,
two directions, i.e. the direction of the propagation and 
the normal direction of the plane coincide.
However, when the shock is rippled with 
the tangential velocity, two pressures lose the balance in the 
tangential direction. This unbalance is expected to be larger
with the increase of the velocity.  
The unstable mechanism indeed works in the ultra relativistic blast wave
as shown in this paper.
%
This result depends on several assumptions and
approximations, but it is quite suggestive.
Suppose, e.g., for the relativistic flow dynamics from $ \Gamma = 10^2$
to 1 associated with the gamma-ray bursts,
that the amplitude increases by more than a factor of $10^{20}$.
The dominant scale of the unstable mode depends
on both the initial disturbance and the growth rate.
If there is no characteristic wavelength in the 
initial stage of the expansion, 
the unstable mode is likely to be 
non-linear at the smaller scale due to the large growth rate.
This mode may lead to a number of fragmentations.
The radiation from the blobs may be reflected by an irregular
time-profile of the gamma-rays.  

In order to check this possibility, further study is necessary.
For dynamical evolution, we must simulate relativistic
flows using axi-symmetric or 3-dimensional numerical codes.
Fluctuations are manifest only in the small 
region after the shock front. The structure may be found
only by using a fine resolution in the numerical calculations.
The unstable mode may be suppressed to a certain value of
the amplitude.
Also, the size and number relations of the fragmentation can be
estimated.
The effect of the flow dynamics will appear in the emission,
but the  gamma-ray emission process is not
clear at moment.
However, the complex structure of the shock front may universally 
give the time history of the burst, irrespective of
the detailed behavior of the emission process.

%
\appendix

\section{Jump Conditions}\label{app.AA}

The difference between the values  on the two
sides of a surface is denoted by $[a] \equiv a_1 - a_2$, 
where the subscript ``$1$'' indicates the fluid 
located before the  shock wave, and 
``$2$'' after the shock wave.
The number density and energy-momentum are
continuous across the shock surface.
The jump conditions for the relativistic adiabatic shock are
given by 
\begin{eqnarray}
 \left[ n\,u^\mu\,{N_s}_\mu \right] &=& 0  \ ,  
\\
 \left[ T^{\mu\,\nu}\,{N_s}_\mu\,{N_s}_\nu \right] &=& 0  \ , 
\\
\left[ T^{\mu\,\nu}\,{N_s}_\mu\,{U_s}_\nu \right] &=& 0  \ , 
\end{eqnarray}
where $T^{\mu\,\nu}$ and  $u^{\mu}$ 
are the energy-momentum tensor and the 4-velocity of the fluid.
The time-like unit vector ${N_s}_\mu$ describes the 
motion of the shock front, and the space-like unit vector 
${U_s}_\mu$ describes the normal direction of the surface.
For a spherically expanding shock wave,
they are given by
$  {N_s}_\mu = \{ \Gamma \,V  , \Gamma ,  0 \} $
and 
$ {U_s}_\mu = \{ \Gamma ,   \Gamma  V , 0 \}$.
We consider linear perturbations of them. The 
explicit forms  are given by
\begin{eqnarray}
  {N_s}_\mu &=& \bigl\{ \Gamma \,\left( V + \Gamma ^2\,\delta V _r \right)
                , \Gamma \,\left( 1 + V \,\Gamma ^2\,\delta V_r \right) 
                , N_{\perp} \bigr\} \ , \\
 {U_s}_\mu &=& \bigl\{ \Gamma \,\left( 1 + V \,{{\Gamma }^2}\,\delta V_r
				 \right)
                , \Gamma \,\left( V + \Gamma ^2\,\delta V _r \right) 
                , U_{\perp} \bigr\} \ ,
\end{eqnarray}
where $\delta V _r$ is the velocity perturbation 
in the radial direction, and 
$ N_{\perp}$ and $  U_{\perp} $ are
the tangential components of the perturbation of the
variables. Their explicit forms  are not necessary in the
calculations.

\section{Analytic Solution} \label{appD}
The coefficients of the analytic solutions
Eqs.~(\ref{ogr-5-last.m3.perturb.eq.1}) --
(\ref{ogr-5-last.m3.perturb.eq.4}) are determined as
\begin{subequations}
\begin{eqnarray}
 a &=& \bigl[28\,\left( k -2  \right) \,
  \bigl\{k\,\left(  4\,s -5   \right)- 12\,s + 7 \bigr\} \bigr.\,\nonumber\\
 &&{}\times\left.
  \left\{
    {k^2}\,\left(  4\,{s^2}+ 9\,s -23  \right)
   - k\,\left( 24\,{s^2} + 50\,s -101 \right)    
    +36\,{s^2} + 69\,s -110 
   \right\}\right]^{-1}  \nonumber\\
   &&{}	\times\left[
	 -8\,{k^5}\,\left(  352\,{s^3}- 584\,{s^2}- 3983\,s  +5905  \right)
	\right. \nonumber\\
     &&{} ~~~ \left.
	  +  {k^4}\,\left( 5616\,{s^4} +  22872\,{s^3}- 115057\,{s^2}
		     - 225180\,s + 425829   \right)
	\right. \nonumber\\
     &&{} ~~~ \left.	   
	 -  {k^3}\,\left(  60624\,{s^4}  +  56344\,{s^3} - 794477\,{s^2}
		           - 518881\,s +1515622    \right)
	\right. \nonumber\\
     &&{} ~~~ \left.	  
	+   {k^2}\,\left( 242352\,{s^4} + 11088\,{s^3} - 2362877\,{s^2}
		       - 277212\,s +  2660875   \right)
	\right. \nonumber\\
     &&{} ~~~ \left.	 
	 - k\,\left( 423792\,{s^4} - 125064\,{s^3} -  3211707\,{s^2}
	               + 457097\,s +2302554  \right)
	\right. \nonumber\\
     &&{} ~~~ \left.	  
	 +  2\,\left(136080\,{s^4} -  61020\,{s^3} - 817893\,{s^2}
		        + 227748\,s  + 392500 \right)
      \right. \nonumber\\
   &&{} ~~~ \left.
	 -3\left\{
	  8\,{k^4}\,\left( 6\,{s^2}- s  -15  \right)
      -{k^3}\,\left(    108\,{s^3}+ 451\,{s^2}- 420\,s   -867   \right)
       \right.\right.\nonumber\\
     &&{} ~~~
      +{k^2}\,\left( 900\,{s^3}+ 1418\,{s^2} - 2409\,s  -2273  \right)\nonumber\\
      &&{} ~~~	-   k\,\left( 2484\,{s^3}+ 1605\,{s^2}  - 4737\,s  -2558\right)
      \nonumber\\
     &&{} ~~~\left.	\left.
	+2268\,{s^3}+ 342\,{s^2} - 3060\,s -1040\right\} f_1 \right]\ ,\\
 b &=&  \bigl[4\,l (l+1)
         \bigl\{k\,\left(  4\,s-5 \right) - 12\,s +7\bigr\}\bigr]^{-1}
          \nonumber\\
       &&{}\times\left[
	    8\,{k^3}\,\left( 20\,{s^2}+ 155\,s -38  \right)
	    - {k^2}\,\left( 384\,{s^3}+ 4012\,{s^2} + 7249\,s  -1265
		         \right)   \right. \nonumber\\
       &&{} ~~~ +  k\,\left( 2304\,{s^3}+ 17844\,{s^2}+ 13577\,s  -1539 
		                       \right)	 \nonumber \\
       &&{} ~~~   -  18\,\left(  192\,{s^3}+ 1208\,{s^2}+ 439\,s -30
		      \right)
	      - \left\{
		  8\,{k^2}\,\left( 2\,s+ 3 \right) \right.\nonumber\\
   && {} ~~~ \left.\left.	 -   k\,\left(  36\,{s^2}+ 125\,s +63 \right)
		 + 2(54\,{s^2}+ 75\,s  + 20 )
		 \right\} f_1  \right] \ ,\\
 c &=&  \left[
	 2\,\bigl\{ 7 - 12\,s + k\,\left( -5 + 4\,s \right)  \bigr\} \,
       	 \left\{ 2\,{k^2}\,\left(  4\,{s^2}+ 23\,s +5  \right)
	  \right.\right. \nonumber \\
	&&{} ~~~ \left.\left.       -   k\,\left( 48\,{s^2}+ 240\,s +43 \right)
	      + 72\,{s^2}+ 306\,s +   46 
	       \right\}    \right]^{-1} \nonumber\\
       && {} \times\left[\left\{
		 8\,{k^2}\,\left(  2\,s+3 \right)
		-   k\,\left(  36\,{s^2}+ 125\,s +63 \right)
		+ 108\,{s^2}+ 150\,s +40 
	      \right\} \right.\nonumber\\
       && {} \left.
	     \times \left\{
		-   2\,{k^2}\,\left( 8\,{s^2} + 34\,s -21 \right)
		+   k\,\left( 96\,{s^2} + 372\,s -149 \right)	       \right.  \right. \nonumber\\
	  &&{} ~~~      - 144\,{s^2} - 504\,s +121
       \bigl.\bigl.
	     + ( 2\,k-3)  f_1     \bigr\}\bigr] ,\\
 d &=&  0.
\end{eqnarray}
\end{subequations}


\end{document}